\documentclass[twocolumn,10pt,journal]{IEEEtran}
\usepackage[left=0.75in,right=0.75in,top=0.75in,bottom=1.5in]{geometry}
\usepackage{amssymb}
\usepackage{amsmath}
\usepackage{textcomp}
\usepackage{amsfonts}
\usepackage{amsthm}
\usepackage{cite}
\usepackage{pdflscape}
\usepackage{tabularx}
\usepackage{paralist}
\usepackage{algorithm}
\usepackage[noend]{algpseudocode}

\usepackage{enumitem}
\usepackage{pdflscape}
\usepackage{multicol}
\usepackage{multirow}
\usepackage{subcaption}
\usepackage{url}
\usepackage{graphicx}
\graphicspath{ {Images/} }
\ifCLASSINFOpdf
\else
 \fi

\begin{document}

\pagenumbering{gobble}

\title{Exploration of TCP Parameters for Enhanced Performance in a Datacenter Environment}

\author{\IEEEauthorblockN{Mohsin Khalil$^*$, Farid Ullah Khan$^\#$
}\\
\IEEEauthorblockA{$^*$National University of Sciences and Technology, Islamabad, Pakistan \\ $^\#$Air University, Islamabad, Pakistan \\
}
}

\maketitle

\begin{abstract}

TCP parameters in most of the operating systems are optimized for generic home and office environments having low latencies in their internal networks. However, this arrangement invariably results in a compromized network performance when same parameters are straight away slotted into a datacenter environment. We identify key TCP parameters that have a major impact on network performance based on the study and comparative analysis of home and datacenter environments. We discuss certain criteria for the modification of TCP parameters and support our analysis with simulation results that validate the performance improvement in case of datacenter environment.

\end{abstract}


\section{Introduction}

Today, majority of the networks being used in home and office environments  generally have speeds ranging from 100 Mbps to 1 Gbps. Although the biggest strength of the Internet is the TCP/IP hourglass as it hides network and applications from each other, however it also hides the flaws unfortunately. Appropriate selection of parameters plays a vital role in affecting the network performance \cite{khalil2018review}, because their improper configuration may result in compromized operation, something that has been proven repeatedly during the performance evaluation of networks \cite{gomez2018tcp}. The impact of communication growth on economy may be gauged from the fact that Information and Communication Technology (ICT) resulted in contributing 1\% GDP increase in case of Singapore \cite{khalil2017feasibility}.

Many studies have been carried out to ascertain TCP performance in diverse environments and various improvements have been suggested in this regard \cite{atxutegi2018use,qian2009tcp,balakrishnan1997comparison,
 pacifico2009improving, wang2007fast,hurtig2012enhanced,yao2008empirical}. Pontes et al. \cite{pontes2017assessment} have demonstrated the impact of various TCP parameters by evaluating their adjustment for mobile devices. Moreover, Saha et al. \cite{saha2015first} have carried out evaluation of TCP in office environment at 60 GHz. However, these results are not valid for the case of a datacenter environment as there is no fit-to-all solution. It has been proved that instead of home environment, a datacenter environment is more suitable to be taken as a blueprint for performance enhancement and buffer length adjustments in data acquisition networks \cite{jereczek2015analogues}.

Linux as an operating system is the typical go-to option for network operators due to it being open-source. The parameters for Linux distribution are optimized for any range of networks ranging up to 1 Gbps link speed in general. For this dynamic range, TCP buffer sizes remain the same for all the networks irrespective of their network speeds. These buffer lengths might fulfill network requirements having relatively lesser link capacities, but fall short of being optimum for 10 Gbps networks. On the basis of Pareto's principle, we segregate the key TCP parameters which may be tuned for performance enhancement in datacenters. We have carried out this study for home and datacenter environments separately, and presented that although the pre-tuned TCP parameters provide satisfactory performance in home environments, however, the network performance in datacenters would be compromized in this case. Therefore, we highlight the important TCP parameters that can be modified for performance enhancement in a datacenter environment and present results in this regard.

This paper is divided into five sections.  Section II of the paper highlights the identification of network parameters whose modification would directly impact the network performance. The criteria and evaluation methodology for selecting the appropriate values of network parameters in Linux operating system has been appended in Section III. The results and observations obtained after tuning of these network parameters are summarized in Section IV, followed by conclusion and scope for future work in Section V.

\section{Selection of TCP Parameters for Direct Impact on Performance}
\vspace{0.1cm}
Various parameters were explored to identify the ones which would have a direct effect on network performance. In order to make this paper self-contained, we briefly dilate upon the key parameters having a significant impact on the datacenter network performance in ensuing paragraphs.

\subsection{Selection of Performance Metric}

In our work, throughput was chosen as the metric for assessing the network performance. It is the most widely used benchmark and is also sometimes used interchangeably with link capacity. So it directly refers to the performance of the network in terms of data rate measured in bits per second (bps). Optimum throughput is pivotal to ensure good network performance. It can be measured by using any of the large number of tools available online with varying levels of accuracy. Since we are identifying the TCP keys on the basis of mice-elephant phenomenon, therefore our approach would yield an improved performance instead of an optimal one.

\subsection{Adjustment in TCP Socket Buffers}

Kernels of operating systems allocate some space to hold TCP data before processing in given socket known as buffers. The receiver keeps on storing the data in buffer before application starts processing on that data and continuously tells sender about space in its buffer. If the receiver buffer is full, sender waits  and keeps on storing data in its sending buffer during time of wait. Socket buffers are very critical in terms of affecting the network throughput. If the socket buffer is very large it will result in occupying more memory making the system slow. If it is very less it will increase packet loss which results in decrease in data transfer speeds.

\subsection{TCP Selective Acknowledgements}

If Selective Acknowledgement (SACK) is enabled, SACK messages with duplicate ACK are also sent to sender with original sequence of received packets. Sender can easily understand the situation after looking at duplicate ACK and sequence of SACK messages and repeated duplicate ACKs are not needed in this process. Sender sends exactly same packet which was missing because whole scenario is known by sender.

\subsection{TCP Queue Length}

Maximum value of queue size should be chosen in such a way so that packet drop may be avoided due to unavailability of memory space or a fast transmitting sender. Generally, the default size of this queue length is 1000 packet count for Linux and most of other operating systems. However, the actual queue size at any time depends upon the characteristics of the network. Small value is recommended in case of high network latencies, however high value can be chosen for high speed connections in specific cases to facilitate large data transfers.

\subsection{Size of Maximum Transmission Unit}

Default Maximum Transmission Unit (MTU) in local area networks is generally 1500 bytes, however it may be increased for environments having link speeds higher than 1 Gbps \cite{rfc,borman2014tcp}. Since MTU assists the TCP in determining the maximum frame size in a transmission, so it may be considered as the upper bound to the size of the frame. The increase in the size of frame will result in a decrease in complexity, because it is similar to the truck transportation analogy; if some cargo from is to be transported from one place to another, less number of heavy-trucks would be needed to carry more cargo at a time, whereas, if same cargo is carried by large number of mini-trucks, the toll plazas at roads would require more processing, wait and time duration. Therefore, large frames will result in better data rates compared to smaller ones. It is better to send large frames only if network quality is good and packet loss probability is low. This is because if network is congested, default packet sizes would be preferred in such situations as large frames will severely impact the network performance in this case.

\section{Evaluation Methodology \& Network Scenarios}

Linux was utilized as an operating system in all the corresponding hosts which were spawn as server / client for analysis of network performance and tweaking of TCP parameters. However, the network selection for carrying out the performance analysis was classified into two phases. The tests / trials during the two phases remained the same with the exception of network selection and adjustment of relevant parameters.

\subsection{Phase-I (Home Environment)}

Phase-I of the experimentation comprised of following two different scenarios:

\begin{itemize}
\item The first case was to establish a local area network between two computers and to check the throughput of the network. The same process was then repeated after tweaking of TCP / network parameters.
\item The second case was to analyze the throughput over the home network at relatively low internet speeds. Similar to first case, the activity was carried out separately for default parameters and same process was repeated multiple times with varying values of TCP parameters. Thus the analysis was carried out to obtain results in two networks within home environment: LAN of bandwidth 1 Gbps and broadband / DSL with bandwidth 10 Mbps.
\end{itemize}

\subsection{Phase-II (Datacenter Environment)}

The Phase-II also involved two different cases, which are listed as follows:

\begin{itemize}
\item The first one involved the virtual machine (VM) deployment in Datacenter environment. For this purpose, two VMs were spawn in East America region on Microsoft Azure such that their network interfaces supported 1 Gbps bandwidth. These VMs were selected from the DS-2 Category in Microsoft Azure. Subsequently, throughput of the network was checked separately for default parameters and same process was repeated with varying values of TCP parameters.
\item Similarly, Case-2 also involved the VM deployment in Datacenter environment. However for this case, two VMs were spawn in East Asia region on Microsoft Azure such that their network interfaces supported 10 Gbps bandwidth. The VMs were selected from the Alpha-8 Category. Subsequently, network analysis was carried out for varying values of TCP parameters.
\end{itemize}

\subsection{Linux Keys Resulting in Significant Performance Impact upon Modification}

Proper adjustment in TCP buffer can result in increase in throughput. Every connection is specified by distinct buffer spaces for input and output. Therefore after thorough study, values of various TCP keys in Linux OS were modified to study the impact on network performance. As a result, the following parameters resulted in major impact on network performance upon modification:

\begin{itemize}
  \item \textit{tcp\_moderate\_rcvbuf}: Dynamically updates receiver buffer size for generic network. Unique bounds for a particular network environment
  \item \textit{tcp\_rmem}: TCP receive buffer memory size
  \item \textit{tcp\_wmem}: TCP transmit buffer memory size
  \item \textit{rmem\_max}: TCP maximum receive window
  \item \textit{wmem\_max}: TCP maximum transmit window
  \item \textit{tcp\_sack}: TCP Selective Acknowledgement
  \item \textit{MTU}: Maximum Transmission Unit
  \item \textit{txqueuelen}: Transmission Queue Length
\end{itemize}

\subsection{Criteria for Modification of TCP Parameters}

Miscellaneous criteria were explored on behalf of which optimum values of TCP parameters can be set to ensure maximum throughput for different connection speeds.

\vspace{0.2cm}
\subsubsection{Analyzing the Bandwidth-Delay Product}

The Bandwidth-delay product (BDP) is amount of data measured in bits or bytes. In fact, it is data that has been transmitted in network but did not receive acknowledgement. It is necessary to increase buffer sizes in accordance with bandwidth delay product to increase throughput of network. In case of home and office environments, latency is generally negligible and bandwidth is very large, which may slightly increase its BDP as compared to other scenarios.

\vspace{0.2cm}
\subsubsection{Utilization of Buffer Allocation}

It is well known that round-trip times (RTT) within a Datacenter environment are less than other conventional network scenarios. For this reason, buffer sizes allocated to switches are much higher than BDP. Zhang et al. in \cite{zhang2014sharing} proposed a switch buffer allocation protocol (SAB) for bandwidth sharing in datacenter networks which assesses the buffer sizes of all the switches along a particular path and then adjusts the congestion windows accordingly. They advocate that the full bandwidth utilization in a network may be ensured by allocating the buffer size greater than its BDP value. In this context, we analyzed the networks in home environment as well as datacenter environment. The networks having buffer sizes less than BDP were assessed for performance after increasing their buffer lengths higher than the BDP value.

\vspace{0.2cm}
\subsubsection{Throughput Measurement}

Throughput can be measured by using any of the available tools with varying levels of accuracy. We kept the network in operation with running \textit{IPERF} for shifts of consecutive 10 hours across several dates for recording the instantaneous throughput at every second in addition to the cumulative average to cover maximum portion of the working day for datacenter environment.

\section{Results and Observations}
\vspace{0.1cm}

By default, the minimum value of TCP transmit and receive memory buffer size is 4096 bytes, while their initial values correspond to 16384 and 87380 bytes respectively. We will analyze that how deviation from default values impacts the performance over different networks.

\subsection{Results from Phase-I}

No performance enhancement was observed by modification of aforementioned TCP parameters in Linux kernel. Upon analysis, it was revealed that by default, TCP receive buffer size is 85 KB which equals 699 Kbits. For Case-1 (LAN), BDP is $10$\textsuperscript{9} bps x 0.12 ms = 120 Kbits and it is even lesser for Case-2 because of much lower bandwidth (10 Mbps). So the default buffer size is still higher than the BDP and no more improvement in network performance is possible in both cases of home environment.

\begin{figure}
  \centering
  \begin{minipage}[b]{0.4\textwidth}
    \includegraphics[width=\textwidth]{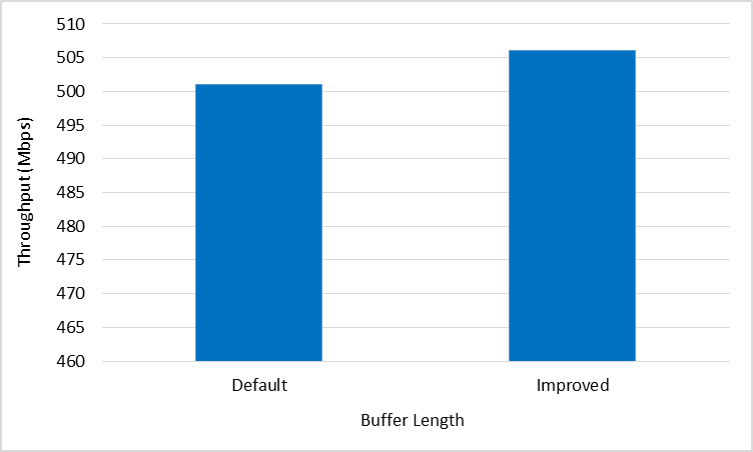}
    \caption{Throughput (in Mbps) vs Buffer size for 1 Gbps link in Datacenter environment}
  \end{minipage}
  \hfill
  \begin{minipage}[b]{0.4\textwidth}
    \includegraphics[width=\textwidth]{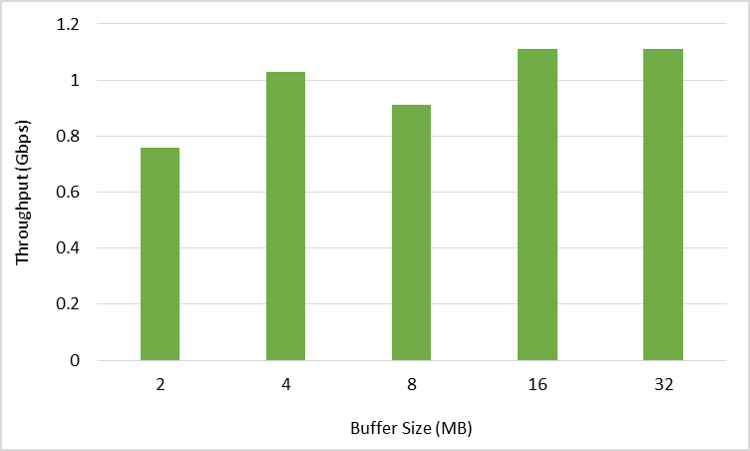}
    \caption{Throughput (in Gbps) vs Buffer size for 10 Gbps link in Datacenter environment}
  \end{minipage}
\end{figure}

\subsection{Results from Phase-II}

In Case-1 (1 Gbps link), small improvement in performance was observed for VMs of Microsoft Azure DS-2 category as a result of buffer size variation. The reason was investigated and it was found out that the BDP for Case-1 is $10$\textsuperscript{9} bps x 1.49 ms = 1.49 Mbits which is higher than the default TCP buffer size of 699 Kbits. So increase in buffer size was carried out and results are appended in Fig. 1. It is pertinent to mention that the throughput was improved from 501 Mbps for default buffer size to 506 Mbps upon increasing buffer size. However, since the link capacity was under 1 Gbps, therefore the modification of other parameters such as MTU and Queue length did not contribute any further improvement in network performance.

On the other hand, the performance improvement was more significant for Case-2 (10 Gbps link) . It is worth highlighting that the BDP for Case-1 was higher than default buffer size, however, the BDP for Case-2 is $10$\textsuperscript{10} bps x 1.58 ms = 15.8 Mbits which is greater than even Case-1 and much higher than default buffer size. So the buffer size has to be increased to maximize link utilization. Same activity was subsequently carried out and the corresponding results are shown in Fig. 2. Throughput (in Gbps) has been plotted against various buffer sizes (in MBs) and higher throughput for high buffer sizes can be seen.

For Case-2 (10 Gbps link), greater MTU size yielded higher throughput. At MTU size of 10,000 bytes, maximum throughput was observed against any buffer length. Fig. 3 depicts the throughput for different MTU sizes against varying buffer lengths in MBs.

Queue length was sufficient for 1 Gbps link in Case-1, but for 10 Gbps it was not enough. So the queue lengths were increased in order to enable high transfer rate. The activity was carried out for queue lengths of 1000 (default), 2000, 4000, 8000 and 16000 packets as shown in Fig. 4. Maximum throughput was observed for 8000 queue length for default MTU size as well as 10,000 byte MTU size.

\begin{figure}
  \centering
  \begin{minipage}[b]{0.38\textwidth}
    \includegraphics[width=\textwidth]{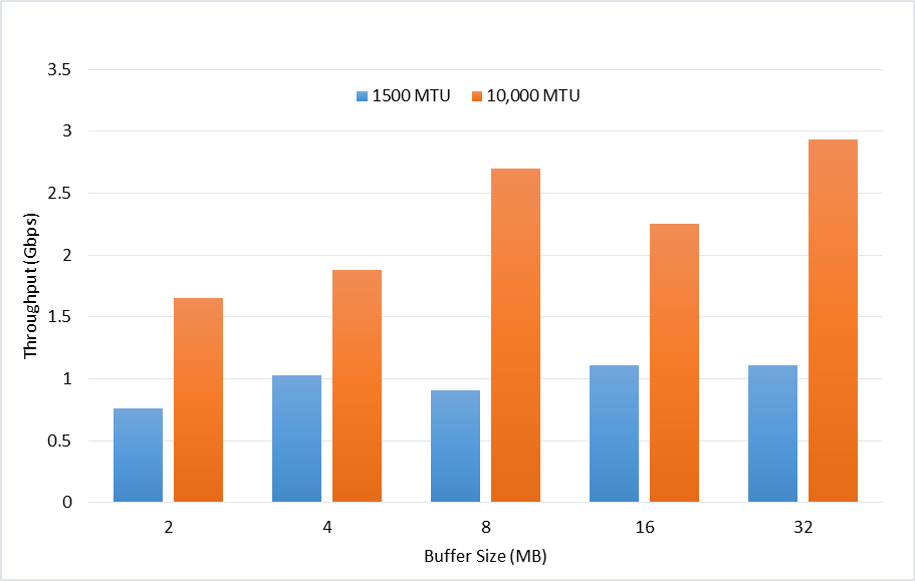}
    \caption{Throughput (in Gbps) for different MTU sizes against buffer lengths (in MBs)}
  \end{minipage}
  \hfill
  \begin{minipage}[b]{0.38\textwidth}
    \includegraphics[width=\textwidth]{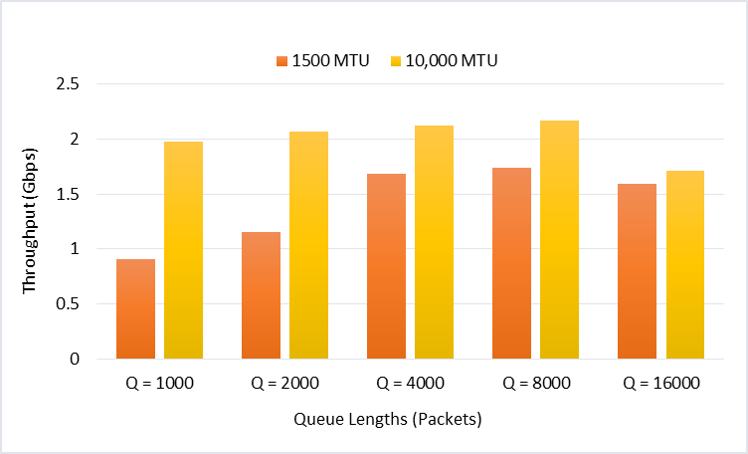}
    \caption{Throughput (in Gbps) for two MTU sizes against varying queue lengths (in packets)}
  \end{minipage}
\end{figure}

\section{Conclusion \& Future Work}

After carrying out performance analysis of various networks, we conclude that for LAN speeds up to 1 Gbps and various other internet connections, performance improvement by modification of TCP parameters is not feasible and the operating system algorithm is smart enough to meet the changing requirements. This is because the default buffer size is higher than the BDP for home environment due to low RTT (around 0.1-0.2 ms), whereas it is higher for broadband case due to low bandwidth. On the other hand, improvement in TCP performance is possible for datacenter environment because even for 1 Gbps network interface, the RTT is at least 10 times higher than that of generic home network which makes BDP high. So for 1 Gbps datacenter case, the performance improvement can be observed by modification of buffer size. However, in case of 10 Gbps datacenter case, the performance improvement is largely contributed by increased buffer size, high MTU and larger queue lengths.

Further extensions of this work may involve the consideration of latency as performance metric, since the networks having frequent speed variation would result in a fluctuating BDP, which would render this approach falling short. Since this work focused on achieving improved performance, a more comprehensive study may be formulated to achieve optimum network performance by consideration of all parameters as constraints with an aim to maximize the network performance for optimum network utilization.

\bibliographystyle{IEEEtran}
\bibliography{Survey_Paper}

\end{document}